\input harvmac
\noblackbox
\newif\ifdraft

\catcode`\@=11
\newif\iffrontpage
\newif\ifxxx
\xxxtrue

\newif\ifad
\adtrue
\adfalse

\parindent0pt

\def\a{\alpha}
\def\b{\beta}
\def\g{\gamma}

\def\s{\sigma}

\def\p{\partial}

\def\gz{\displaystyle{\mathrel{\mathop g^{\scriptscriptstyle{(0)}}}}} 
\def\go{\displaystyle{\mathrel{\mathop g^{\scriptscriptstyle{(1)}}}}} 
\def\gt{\displaystyle{\mathrel{\mathop g^{\scriptscriptstyle{(2)}}}}} 
\def\gth{\displaystyle{\mathrel{\mathop g^{\scriptscriptstyle{(3)}}}}} 
\def\gn{\displaystyle{\mathrel{\mathop g^{\scriptscriptstyle{(n)}}}}} 

\def\na{\nabla}

\parindent0pt

\def\{{\lbrace}
\def\}{\rbrace}

\def\a{\alpha}
\def\b{\beta}
\def\c{\gamma}

\def\s{\sigma}
\def\g{{g}}

\def\p{\partial}

\def\gz{\displaystyle{\mathrel{\mathop g^{\scriptscriptstyle{(0)}}}}} 
\def\go{\displaystyle{\mathrel{\mathop g^{\scriptscriptstyle{(1)}}}}} 
\def\gt{\displaystyle{\mathrel{\mathop g^{\scriptscriptstyle{(2)}}}}} 
\def\gn{\displaystyle{\mathrel{\mathop g^{\scriptscriptstyle{(n)}}}}} 

\def\Box#1{\mathop{\mkern0.5\thinmuskip
           \vbox{\hrule\hbox{\vrule\hskip#1\vrule height#1 width 0pt\vrule}
           \hrule}\mkern0.5\thinmuskip}}
\def\box{\displaystyle{\Box{7pt}}}

\def\Box#1{\mathop{\mkern0.5\thinmuskip
           \vbox{\hrule\hbox{\vrule\hskip#1\vrule height#1 width 0pt\vrule}
           \hrule}\mkern0.5\thinmuskip}}

\def\{{\lbrace}
\def\}{\rbrace}

\def\a{\alpha}
\def\b{\beta}
\def\c{\gamma}

\def\s{\sigma}

\def\p{\partial}

\def\na{\nabla}

\def\dal{\displaystyle{\Box{7pt}}}
\def\frac#1#2{{\scriptstyle{#1}\over\scriptstyle{#2}}}

\def\abstract#1{
\vskip.5in\vfil\centerline
{\bf Abstract}\penalty1000
{{\smallskip\ifx\answ\bigans\leftskip 2pc \rightskip 2pc
\else\leftskip 5pc \rightskip 5pc\fi
\noindent\abstractfont \baselineskip=12pt
{#1} \smallskip}}
\penalty-1000}


\lref\AdSreview{
O.~Aharony, S.~S.~Gubser, J.~M.~Maldacena, H.~Ooguri and Y.~Oz,
``Large N field theories, string theory and gravity,''
Phys.\ Rept.\  {\bf 323}, 183 (2000), hep-th/9905111.}

\lref\WittenAdS{
E.~Witten,
``Anti-de Sitter space and holography,''
Adv.\ Theor.\ Math.\ Phys.\  {\bf 2}, 253 (1998), hep-th/9802150.}

\lref\Duff{M.~J.~Duff,
``Twenty years of the Weyl anomaly,''
Class.\ Quant.\ Grav.\  {\bf 11}, 1387 (1994), hep-th/9308075
and references therein.}

\lref\dHSS{S.~de Haro, S.~N.~Solodukhin and K.~Skenderis,
``Holographic reconstruction of spacetime and renormalization in the
AdS/CFT correspondence,'' Commun.\ Math.\ Phys.\  {\bf 217}, 595 (2001),
hep-th/0002230.}

\lref\BFS{M.~Bianchi, D.~Z.~Freedman and K.~Skenderis,
``How to go with an RG flow,''
JHEP {\bf 0108}, 041 (2001), hep-th/0105276;
``Holographic renormalization,''
Nucl.\ Phys.\ B {\bf 631}, 159 (2002), hep-th/0112119.}

\lref\Skenderis{K. Skenderis,
``Lecture notes on holographic renormalization,''
Class.\ Quant.\ Grav.\  {\bf 19}, 5849 (2002), hep-th/0209067.}

\lref\DS{S.~Deser and A.~Schwimmer,
``Geometric Classification of Conformal Anomalies in Arbitrary Dimensions,''
Phys.\ Lett.\ B {\bf 309}, 279 (1993), hep-th/9302047.}

\lref\ISTY{C.~Imbimbo, A.~Schwimmer, S.~Theisen and S.~Yankielowicz,
``Diffeomorphisms and holographic anomalies,''
Class.\ Quant.\ Grav.\  {\bf 17}, 1129 (2000),
hep-th/9910267.}

\lref\Penrose{R.Penrose and W. Rindler, ``Spinors and
Spacetime,'' CUP 1986, vol.2, chapter 9.}

\lref\BH{J.D. Brown and M. Henneaux, ``Central Charges in the
Canonical Realization of Asymptotic Symmetries: an Example from
Three-Dimensional Gravity,'' Commun. Math. Phys. {\bf 104} (1986)
207.}

\lref\BNTnote{M. Blau, K. Narain and G. Thompson, On the FG Expansion
to ${\cal O}(h)$, private communication}

\lref\FG{C.~Fefferman and R.~Graham, ``Conformal Invariants,''
Ast\`erisque, hors s\'erie, 1995, p.95.}

\lref\APTY{O.~Aharony , J.~Pawelczyk , S.~Theisen and S.~Yankielowicz ,
``A note on anomalies in the ADS/CFT correspondence ,'' Phys.\ Rev. D60
(1999)066001 ,hep-th/9901134 }

\lref\ST{ A.~Schwimmer and S.~Theisen, ``Diffeomorphisms,
anomalies and the Fefferman-Graham ambiguity,'' JHEP {\bf 0008},
032 (2000), hep-th/0008082.}

\lref\HS{M.~Henningson and K.~Skenderis,
``The holographic Weyl anomaly,'' JHEP {\bf 9807}, 023 (1998), hep-th/9806087.}

\lref\NO{S.~Nojiri and S.~D.~Odintsov,
``On the conformal anomaly from higher derivative gravity in AdS/CFT
correspondence,'' Int.\ J.\ Mod.\ Phys.\ A {\bf 15}, 413 (2000)
hep-th/9903033.}

\lref\BNTone{
M.~Blau, K.~S.~Narain and E.~Gava,
``On subleading contributions to the AdS/CFT trace anomaly,''
JHEP {\bf 9909}, 018 (1999), hep-th/9904179.}

\lref\Fulling{S.A. Fulling, R.C. King, B.G. Wybourne and C.J. Cummings,
``Normal forms for tensor polynomials: I. The Riemann tensor'',
Class. Quantum Grav. {\bf 9} (1992) 1151.}

\lref\AS{M Abramowitz and I. Stegun, ``Handbook of Mathematical Functions,''
National Bureau of Standards Applied Mathematics Series 55,
tenth printing 1972.}

\lref\dBVV{
J.~de Boer, E.~Verlinde and H.~Verlinde,
``On the holographic renormalization group,''
JHEP {\bf 0008}, 003 (2000), hep-th/9912012.}

\lref\KMM{J.~Kalkkinen, D.~Martelli and W.~M\"uck,
``Holographic renormalisation and anomalies,''
JHEP {\bf 0104}, 036 (2001), hep-th/0103111.}

\lref\MM{
D.~Martelli and W.~M\"uck,
``Holographic renormalization and Ward identities with the
Hamilton-Jacobi method,''
Nucl.\ Phys.\ B {\bf 654}, 248 (2003), hep-th/0205061.}

\lref\AF{G.~Arutyunov and S.~Frolov, ``Three-point Green function of the
stress-energy tensor in the AdS/CFT correspondence,'' Phys.\ Rev.\ D{\bf 60},
 026044 (1999), hep-th/9912210.}

\lref\Erdmenger{J.~Erdmenger,
``Conformally covariant differential operators: Properties and  applications,''
Class.\ Quant.\ Grav.\  {\bf 14}, 2061 (1997), hep-th/9704108.}

\lref\Tup{S. Theisen, unpublished.}


\Title{\vbox{ \rightline{\vbox{\baselineskip12pt
\hbox{AEI-2003-075} \hbox{hep-th/0309064}}}}}
{Universal Features of Holographic
Anomalies\footnote{$^{\scriptscriptstyle*}$}{\sevenrm Partially
supported by GIF, the German-Israeli Foundation for Scientific
Research, the Alexander-von-Humboldt Foundation, 
the European Commission RTN programmes
HPRN-CT-2000-00131 and HPRN-CT-2000-00122, 
Minerva Foundation, the
Center for Basic Interactions of the Israeli Academy of Sciences and
the Einstein Center.}}
\vskip 0.3cm
\centerline{ A.~Schwimmer$^a$ and S.~Theisen$^b$ } 
\vskip 0.6cm
\centerline{$^a$ \it Department of Physics of Complex Systems,
Weizmann Institute, Rehovot 76100, Israel} 
\vskip.2cm
\centerline{$^b$ \it Max-Planck-Institut f\"ur Gravitationsphysik,
Albert-Einstein-Institut, 14476 Golm, Germany} 
\vskip 0.0cm

\abstract{We study the mechanism by which gravitational actions reproduce the
trace anomalies of the holographically related conformal field theories.
Two universal features emerge:
a) the ratios  of type B trace anomalies  in any even dimension
are independent of the gravitational
action being uniquely determined  by the underlying algebraic structure
b) the normalization of the type A and the overall normalization
of the type B anomalies are given by action dependent expressions with the
dimension dependence completely fixed.}

\Date{\vbox{\hbox{\sl {September 2003}}
}}
\goodbreak

\parskip=4pt plus 15pt minus 1pt
\baselineskip=15pt plus 2pt minus 1pt

\noblackbox

\newsec{Introduction}

The calculation of trace anomalies \Duff\
provides a remarkable test \WittenAdS\HS\ of the
AdS/CFT correspondence \AdSreview .
Besides its very interesting result the calculation
indicated a new, highly nontrivial mechanism by which an anomaly can appear
in an essentially classical setup.

The algebraic structure underlying the anomaly calculation was
studied in \ISTY. The Weyl transformation on the boundary CFT is
embedded in a subgroup of the diffeomorphisms acting on the odd
dimensional gravitational action (``PBH transformation'' after
Penrose \Penrose\ and Brown and Henneaux \BH). The PBH
transformations act on a general solution $g_{ij}(x,\rho)$ of the
equations of motion in a nonlinear fashion, constraining its form.

Though the functional dependence of $g_{ij}$ is not completely determined
one can isolate its part relevant for the anomaly calculation which is strongly
constrained by the PBH transformations.
These terms are
singled out by a  cohomological structure which was studied in
\ST. In the present paper we  study in detail the relation between
the cohomologically nontrivial part of $g_{ij}$ and the trace anomalies.

We show that the anomalies are related to the relevant part of
$g_{ij}$  linearly . The coefficients entering the relation depend
on the gravitational action  but have a universal dependence on
the dimension. This relation provides a rationale for the
existence of a nontrivial cohomology for $g_{ij}$ and implicitly
for the Fefferman-Graham(``FG'') ambiguity \FG,\ST. Then the
constraints imposed by the PBH transformation on $g_{ij}$ get
translated into relations between trace anomalies in the same
dimension and different dimensions for a fixed gravitational
action.

By studying the PBH transformation for the cohomologically non
trivial part of $g_{ij}$ we  conclude that (i) the overall
normalization of type A and B anomalies \DS\ are gravitational
action dependent but the dimension dependence is universal; (ii)
the ratios between the terms responsible for the various type B
anomalies are completely fixed.

In  Section 2
we discuss the general relation between trace anomalies and the
cohomologically nontrivial part of  $g_{ij}$.
We use dimensional regularization which provides unique signatures
for the  two quantities allowing us to relate them linearly .

In Section 3 we calculate the exact expression implied by the PBH
transformations for $g_{ij}$ expanded to first order in curvature. We
interpret this result as giving a unique relation between certain type B
terms in all dimensions.

In Section 4 we calculate exactly using the PBH transformations all the
type B terms relevant in $d=6$ and we show that their coefficients are
completely fixed.
In conjunction with the results of Section 3 this indicates that
all the coefficients of type B terms in all dimensions are fixed by the
PBH transformation.

In Section 5 we check the universal results for the anomalies against
the standard calculation for a  gravitational action containing
arbitrary  terms quadratic in the curvatures.

In Section 6 we summarize our results and discuss the implications for
the general structure of trace anomalies in conformal theories.

In an Appendix we review the relevant features of PBH transformations.

\newsec {The relation between the cohomology of $g_{ij}$ and trace anomalies}

We review first  the well understood signal for  trace anomalies
in dimensional regularization \DS.

We start with the type A anomaly for which there is no true
divergence in $d=2n$. As a consequence the effective action which
is Weyl invariant away from $2n$  dimensions can be decomposed
into two pieces:
\eqn\typeA{W_d(g) = W_d^{({\rm nl})} + {\mu^{d-2n}\over d-2n} \int
d^d x \sqrt g E_{2n}(g)  }
where $E_{2n}$ is the $2n$ dimensional Euler density and $\mu$ is a mass
scale.

The first nonlocal term has  a finite limit for $d=2n$  and the second one
has limit $0$: in dimensional regularization the special relations
valid in integer dimensions are implemented first.

The Weyl variation of the action  in $d=2n$ can be calculated as
the variation (with negative sign) in $d$ dimensions of the
second, local term and it is proportional to $E_{2n}$.

For the type B anomaly the effective action in $ d$ dimensions has
the generic form:
\eqn\typeB{W_d(g)={1\over d-2n}\int d^dx \sqrt g C\dots \box^{-d/2 +n}\dots C
-{\mu^{d-2n}\over d-2n}\int d^d x\sqrt g C\dots C  }
where we denoted symbolically by $C\dots C$ a local expression which transforms
under Weyl rescalings in a homogenous fashion with weight $2n$.

In this case
the first term is Weyl invariant in $ d$-dimensions and has a genuine
ultraviolet divergence represented by the explicit pole term. In order to
have a well defined limit in $d=2n$ we need a local
counter term which is the second term in \typeB\ breaking explicitly
the Weyl invariance. The Weyl variation which is finite
comes now from the second term and  gives an expression proportional
to $C\dots C$.

We see that both the type A and type B anomalies are
finally given (with negative signs) by variations in $d$ dimensions
of the local expressions represented by the second terms in the r.h.s. of
\typeA, \typeB. We remark that exactly in $d=2n$ the dependence on
the scale $\mu$ disappears
such that the anomaly does not violate global dilation invariance.

The above mechanism has an exact counterpart in the holographic
context.
The gravitational action evaluated on a solution of the equations
of motions with boundary value $\gz_{ij}$ is invariant under
diffeomorphisms which include as a subgroup the PBH
transformations and therefore under Weyl transformations on the
boundary. The potential anomalous violation arises due to the
integration over $\rho$ which is potentially divergent at
$\rho=0$. This infrared divergence replaces the ultraviolet
divergence in the conformal field theory.

A calculation involving an exact
integration over $\rho$ between $0$ and $\infty$ would produce in dimensional
regularization the invariant terms in \typeA, \typeB. Alternatively one
can produce directly the local terms from which the anomaly can be
obtained following the procedure described above. For this one uses
an expansion of the solution in integer powers of $\rho$ multiplying
local expressions of $\gz$ (the Fefferman-Graham expansion).

In addition the integration on
$\rho$ is limited between $0$ and $\bar\rho$. Now the integration over
$\rho$ is explicit and in dimensional regularization  $\rho=0$ does not
contribute. Therefore around $d=2n$ one gets terms
\eqn\loc{\Delta W=\sum {\bar\rho^{{1\over2}(2n-d)} \over 2n-d}
\int d^d x \sqrt{\gz(x)}\, b_n(x)}
where $b_n$ are either $E_{2n}$ for type A or one of the
expressions transforming
homogenously which we denoted by $C\dots C$ for type B. Obviously $\bar\rho$
plays the role of the mass scale $\mu^{-2}$ and the variations of
the local terms (i.e. anomalies) become $\bar\rho$- independent for $d=2n$.

The above calculation, being classical, allows, however an alternative path:
derivatives of the action with respect to the initial conditions $\gz$
(``the energy momentum tensor'')
reduce to boundary terms in the usual Hamilton-Jacobi manner
since the action is evaluated on a classical solution\foot{The Hamilton-Jacobi
approach was used in the holographic context in \dBVV\KMM\MM.}.
Since the derivatives of \loc\
with respect to $\gz$ have explicit poles they still carry the complete
information about the anomalies. It follows that the boundary terms
which are local expressions in terms of $g_{ij}(x,\bar\rho)$ should
have the same poles and should carry directly the information about anomalies.

In dimensional regularization the contribution of the boundary $\rho=0$
being put to $0$ the whole contribution will come from an expression involving
$g_{ij}(x,\rho)$ evaluated at $\rho=\bar\rho$.
Indeed in \ST\
poles were shown to appear in the Feffermann-Graham expansion of
$g_{ij}(x,\rho)$
as a consequence of the existence  of a nontrivial cohomology involving
the PBH transformations. The non trivial classes
are in one to one correspondence with the derivatives of the corresponding
local anomaly terms:

there is a unique type A class  for each even dimension $2n$:
\eqn\typeAc{A^E_{(n)ij}= {1\over d-2n}{1\over \sqrt g }{\delta \over \delta
g_{ij}} \int d^d x \sqrt g E_{2n}}
and several type B classes (``Bach tensors'') corresponding to the
type B anomalies:
\eqn\typeBc{A^B_{(n)ij}= {1\over d-2n}{1\over \sqrt g }{\delta
\over \delta g_{ij}} \int d^d x \sqrt{g}C\dots C}
Their number depends on the total
number of derivatives acting on the metric: there
is one of order four, three of order six, etc.
We note that $g^{ij}A^E_{(n)ij}={1\over2}E_{2n}$ and
$g^{ij}A_{(n)ij}^B={1\over2} C\dots C$.

Obviously, since the exact form of the boundary terms depends on the
gravitational action the explicit relation between the trace anomalies and the
non trivial cohomological classes \typeAc,\typeBc\ will also depend on the
action. We illustrate in detail this relation for the case of the simplest
action which has an AdS solution:
\eqn\actiona{S=\int d^dx d\rho \sqrt{G}(\hat R(G)-2\Lambda)}
where $\Lambda={1\over2}d(d-1)$.

With the FG ansatz (A.1) for the metric one finds
\eqn\expansion{\eqalign{
\sqrt{G}&={1\over2}\rho^{-1-d/2}\sqrt{g(\rho)}\cr
\hat R&=d(d+1)+\rho R-2(d-1)\rho g^{ij}g'_{ij}
-3\rho^2 g^{ij}g^{kl}g'_{ik}g'_{jl}
+4\rho^2 g^{ij}g''_{ij}+\rho^2(g^{ij}g'_{ij})^2}}
In the second line and, until further notice, below,
all quantities are computed with $g(\rho)$. Inserting this into the action
gives
\eqn\actionb{\eqalign{
S&={1\over2}\int d^d xd\rho\rho^{-1-d/2}\sqrt{g}\left\lbrace
\rho R-2(d-1)\rho g^{ij}\p_\rho g_{ij}
-3\rho^2 g^{ij}g^{kl}\p_\rho g_{ik}\p_\rho g_{jl}\right.\cr
&\qquad\qquad\left.+4\rho^2 g^{ij}\p_\rho^2 g_{ij}
+\rho^2(g^{ij}\p_\rho g_{ij})^2\right\rbrace}}
Varying the action w.r.t. to $g_{ij}$ one obtains
\foot{We use the following notation:
${\rm tr}g'=g^{ij}g'_{ij},\, g'^{ij}=g^{ik}g^{jl}g'_{kl}$, etc.}
\eqn\variation{\eqalign{ \delta S&={1\over2}\int d^d x
d\rho\sqrt{g}\rho^{-1-{d\over2}}\left\lbrace {1\over2}\rho
Rg^{ij}-\rho R^{ij}-2\rho^2 (g'')^{ij} -\rho^2(\tr g')(g')^{ij}
-{3\over2}\rho^2\tr(g'^2)g^{ij}\right.\cr
&\left.-2\left(1-{d\over2}\right)\rho(g')^{ij} +2\rho^2
(g'^2)^{ij}+2\left(1-{d\over2}\right)\rho(\tr g')g^{ij}
+2\rho^2\tr(g'')g^{ij}+{1\over2}(\tr g')^2 g^{ij}\right\rbrace
\delta g_{ij}+{\rm b.t.}}}
where the boundary terms ($b.t.$) arise from the integrations by parts
w.r.t. $\rho$.
The expression which multiplies $\delta g_{ij}$
are precisely the $(ij)$ components of the Einstein equations
which follow from the action \actiona.
The boundary terms are
\eqn\bt{
{\rm b.t.}=\int d^d x\sqrt{g}\left.\left\lbrace
2\rho^{1-{d\over2}}g^{ij}\delta g'_{ij}-\rho^{-{d\over2}}g^{ij}\delta g_{ij}
-\rho^{1-{d\over2}}(g')^{ij}\delta g_{ij}\right\rbrace
\right|_{\rho=0}^{\rho=\bar\rho}}
Note that through the solution of the PBH equations all terms in
the $\rho$-expansion of $g(x,\rho)$ are functions of
${\displaystyle \gz_{ij}(x)}$. The energy momentum tensor is then
the functional derivative of \bt\ w.r.t. ${\displaystyle
\gz_{ij}}$. In order to isolate the characteristic leading poles
in $d-2n$ with the accompanying $\bar\rho^{d/2-n}$  powers in the
functional derivative of \bt\ we should pick terms where one of
the factors is ${\displaystyle\gn_{ij}}$, the coefficient of $\rho^n$
in the expansion of $g_{ij}(x,\rho)$ around $\rho=0$, all the
others being $\gz_{ij}$. After taking into account that the trace
of $\gn_{ij}(x)$ does not have poles at $d=2n$ and therefore it is
cohomologically trivial, we obtain for the term having the pole at
$d=2n$:
\eqn\tone{ {1\over\sqrt{\gz}}{\delta\over\delta\gz^{ij}}~
{\rm b.t.}= -n\g_{ij}^{(n)}\bar\rho^{n-d/2}+
\hbox{cohomologically trivial}}
in agreement with the results of \dHSS.

We remark that this way of doing the calculation is insensitive to the
presence of explicit nonsingular boundary terms at $\rho=0$.
It can also be straightforwardly applied to more general
gravitational actions, as we will now demonstrate.

An example which we will need in the following and which
appears as a gravity dual of ${\cal N}=2$ super-conformal field
theories in four dimensions \NO,\BNTone\ is the
gravitational lagrangian containing general quadratic terms in the
curvatures:
\eqn\lag{{\cal L}=\hat R-2\Lambda+\a\hat R^2+\b\hat R_{\mu\nu}\hat R^{\mu\nu}
+\gamma\hat R_{\mu\nu\rho\sigma}\hat R^{\mu\nu\rho\sigma}}
Repeating the steps outlined above we obtain for the term having
the pole at $d=2n=4$ and the dependence $\bar\rho^0$:
\eqn\tonee{
{1\over\sqrt{\gz}}{\delta\over\delta\gz{}^{ij}}~{\rm b.t.}=
-2(1+40 \alpha+8 \beta -4 \gamma) \gt_{ij}
+\hbox{coho trivial}}
{}From equations \tone,\tonee\ the anomalies can be identified; once we
isolate the pole terms the anomalies  can be obtained also by taking the trace
of these terms.

\newsec{The solution of the PBH equations to leading order in the curvatures}

In this section we start a systematic study of the PBH equations.
The conclusion will be that all the type B cohomologically
nontrivial contributions are uniquely determined to all orders in
$\rho$.

We work to first order in the curvature. To start with we allow
the most general covariant expression linear in the
curvatures\foot{Note that each derivative $\nabla_i$ is
accompanied by a factor $\sqrt{\rho}$.}
\eqn\ansatz{
g_{ij}(\rho)=g_{ij}+\alpha(\rho\dal)\rho R_{ij}+\beta(\rho\dal)\rho g_{ij}R
+\gamma(\rho\dal)\rho^2\nabla_i\nabla_j R+{\cal O}(R^2)}
Here all quantities
($R,\,\nabla,$ etc.) are with respect to the metric on the boundary,

$\displaystyle{g_{ij}\equiv\gz_{ij}}$,
except where the $\rho$-dependence is explicitly given. We always suppress
the $x$-dependence.

We calculate first the curvature independent piece in the Weyl transformation
of \ansatz :
\eqn\ansatztrans{
\eqalign{
\delta g_{ij}(\rho)&=2\s g_{ij}+\alpha(t)\rho[(d-2)\nabla_i\nabla_j\s
+g_{ij}\dal\s]\cr
&\qquad+\beta(t)\rho\g_{ij}[2(d-1)\dal\s]
+\gamma(t)\rho^2[2(d-2)\nabla_i\nabla_j\dal\s]}
}
where we have defined $t=\rho\dal$.
On the other hand, we can also expand the r.h.s. of eq.(A.2)
to calculate the curvature independent piece and  we find
\eqn\transform{
\delta g_{ij}(\rho)=2\s g_{ij}+\rho\nabla_i\nabla_j\s
}
Comparison of \ansatztrans\ and \transform\ gives
\eqn\recursion{
\eqalign{ t\alpha(t)+2(d-1)t\beta(t)&=0\cr
(d-2)\alpha(t)+2(d-1)t\gamma(t)&=1}}
If we define
\eqn\defbetabar{
t\bar\beta=1+2(d-1)(d-2)\beta}
we get
\eqn\abc{\eqalign{
\beta(t)&=-{1\over2(d-1)(d-2)}+{\bar\beta t\over 2(d-1)(d-2)}\cr
\alpha(t)&={1\over(d-2)}-{\bar\beta t\over(d-2)}\cr
\gamma(t)&={\bar\beta\over 2(d-1)}}}
Inserting this into \ansatz\ we obtain
\eqn\varg{\eqalign{
g(\rho)_{ij}&=g_{ij}+\rho\go_{ij}+\rho^2\bar\beta(t)
\left\lbrace{1\over2(d-1)}\nabla_i\nabla_j R+{1\over2(d-1)(d-2)}g_{ij}\dal R
-{1\over(d-2)}\dal R_{ij}\right\rbrace+{\cal O}(R^2)\cr
&\equiv g_{ij}+\rho\go_{ij}+\rho^2\bar\beta(t)X_{ij}+{\cal O}(R^2)}}
where, unlike the $\gn_{ij}$ for $n>1$,
$\go_{ij}$ is uniquely fixed by (A.2) \ISTY\ and is
\eqn\gone{\go_{ij}={1\over d-2}\left(R_{ij}-{1\over2(d-1)}g_{ij}R\right)\,.}
We remark that $X_{ij}$ is simply
\eqn\coc{ X_{ij}=-{1\over d-3} \nabla^k \nabla^l C_{ikjl}}
and therefore the terms appearing in \varg\ belong to the type B
cohomologically nontrivial Bach tensors generated by
$\int d^d x  \sqrt {\gz} C^{ijkl} \box^n C_{ijkl} $ in dimension
$d=2n+4$.

Some basic properties of $X_{ij}$, which will be used below, are
\eqn\Xprop{\eqalign{
X_i{}^i&=0\cr
\nabla^i X_{ij}&=0+{\cal O}(R^2)}}
What remains is to determine $\bar\beta(t)$.
We will do this by comparing the Weyl variation of \varg\ with
the expansion of eq.(A.2) to first order in the curvature.
We start with the latter.
It will be sufficient to work
to first order in $\nabla_i\s$. This becomes clear once one
realizes that the Weyl variations of the ${\cal O}(R^2)$ terms never
generate any terms which are linear in $R$ and with only one derivative
acting on $\sigma$. In the computation one has to choose a basis for
the possible terms. The basis we choose is that we always move
all $\dal$'s to the left of explicit $\nabla_i$'s. One then has to
use the explicit expression for
$[\nabla_j\nabla_j,\dal]\s$ which produces terms
${\cal O}(R,\nabla_i\s)$.
If we write
\eqn\gZ{
g_{ij}(\rho)=g_{ij}+Z_{ij}(\rho)+{\cal O}(R^2)}
we obtain
\eqn\ginv{
g^{ij}(\rho)=g^{ij}-Z^{ij}(\rho)+{\cal O}(R^2)\,\qquad
Z^{ij}(\rho)=g^{ik}g^{jl}Z_{kl}(\rho')}
To this order (A.3) is
\eqn\ai{\eqalign{ a^i(\rho)&={1\over2}\int_0^\rho d\rho'
g^{ij}(x,\rho')\partial_j\s\cr &={1\over2}\int_0^\rho
d\rho'\left\lbrace g^{ij}-Z^{ij}(\rho) \right\rbrace\partial_j\s}}
Eq.(A.2) also involves terms $g_{jk}(\rho)\nabla_i a^k(\rho)$.
They must also be expanded to ${\cal O}(R)$. Doing all of this we
find
\eqn\twonine{ \delta g_{ij}(\rho)=-{1\over2}\int_0^\rho d\rho'
\left\lbrace\nabla_i Z_j{}^l(\rho')+\nabla_j
Z_i{}^l(\rho')\right\rbrace \nabla_l\s+{1\over2}\rho\nabla_l
Z_{ij}(\rho)\nabla^l\s}
Next, expand
\eqn\Zexp{\eqalign{
Z_{ij}(\rho)&=\rho\go_{ij}+\rho^2\bar\beta(\rho\dal)X_{ij}\cr
\bar\beta(\rho\dal)&=\sum_{n=0}^\infty\beta_n \rho^n\dal^n}}
Inserting this into \twonine\
we find
\eqn\deltagone{
\delta g_{ij}(\rho)={1\over2}\sum_{n=0}^\infty
\beta_{n-1}\rho^{n+2}\dal^{n-1}\left\lbrace\nabla_kX_{ij}
-{1\over (n+2)}\left(\nabla_i X_{jk}+\nabla_j X_{ik}\right)
\right\rbrace\nabla^k\s}
This expression for $\delta g_{ij}(\rho)$, which is valid to
${\cal O}(R,\nabla\sigma)$, but to all orders in $\rho$, is our
first result.

Next we compute $\delta(\bar\beta(\rho\dal)X_{ij})$.
Here the main result is
\eqn\deltaX{
\delta(\dal^n X_{ij})=-(n+2)(2n+4-d)\dal^{n-1}\left\lbrace
\nabla^k X_{ij}-{1\over (n+2)}
(\nabla_i X_{j}{}^k+\nabla_j X_{i}{}^k)\right\rbrace\nabla_k\s}
To derive it eq.\Xprop\ is used.
Some intermediate results are
\eqn\imtermediate{\eqalign{
\delta(\dal^n X_{ij})&=\dal^n\delta X_{ij}
+\dal^{n-1}\left\lbrace n(d-4-2n)\nabla_k X_{ij}
+2n(\nabla_i X_{jk}+\nabla_j X_{ik})\right\rbrace
\nabla^k\s\cr
\dal^n\delta X_{ij}&=\dal^{n-1}\left\lbrace 2(d-4-2n)\nabla_k X_{ij}
-(d-4)(\nabla_i X_{jk}+\nabla_j X_{ik})\right\rbrace\nabla^k\s}}
Again, all calculations are to ${\cal O}(R,\nabla\s)$ in the
basis where all $\dal$'s are moved all the way to the left.

Using \deltaX\ together with eq.\varg\ we obtain
\eqn\deltagtwo{\delta g(\rho)_{ij}=
-\sum_{n=0}^\infty \rho^{n+2}\beta_n (n+2)(2n+4-d)\dal^{n-1}
\left\lbrace\nabla^k X_{ij}-{1\over(n+2)}(\nabla_i X_j{}^k+\nabla_j X_j{}^k)
\right\rbrace\nabla_k\s}
Comparison of \deltagone\ and \deltagtwo\ gives
\eqn\comparison{
\beta_n=-{1\over 2(n+2)(2n+4-d)}\beta_{n-1}}
with $\beta_0=-{1\over 4(d-4)}$.
Solving the recursion relation we finally get
\eqn\final{\eqalign{
g(\rho)_{ij}&=g_{ij}+\rho\go_{ij}
-\sum_{n=0}^\infty\rho^{n+2}{2\over2^{(n+2)}}{1\over(n+2)!}
{1\over(d-4)(d-6)\cdots(d-2(n+2))}\dal^n X_{ij}+{\cal O}(R^2)\cr
&=g_{ij}+\rho\go_{ij}-2(d-2)\rho^2\sum_{n=0}^{\infty}
{1\over 2^{2(n+2)}}{1\over(n+2)!}
{\Gamma({1\over2}d-(n+2))\over\Gamma({d\over2})}\rho^{n}\dal^n X_{ij}
+{\cal O}(R^2)}}
Therefore all the terms in the local FG expansion are uniquely determined
to this order. This is a consequence of the fact that there are no
terms linear in the curvature transforming homogenously whose coefficient
would be undetermined. The terms have poles in even dimensions signaling their
cohomologically nontrivial nature. The component $\gn$ has a leading
pole at $d=2n$ and poles at all the lower even dimensions. While these
secondary poles should be present in $g_{ij}(x,\rho)$ in accordance
with the FG ambiguity, they should not give rise to anomalies : $\rho$
gives the correct scale dependence for $g_{ij}(x,\rho)$ at the pole but
for an anomaly one would need a negative power dependence on
the  scale $\mu$ in contradiction with the analytical structure of CFT.

The knowledge of all the local terms in \final\ allows the calculation
of all the anomalies to this order in the curvture,
the main purpose of the present
paper. Having an exact solution enables us, however, to study as a
byproduct the structure of the FG expansion and in  particular
the FG ambiguity on an all order in $\rho$ expression.

The recursion relation \comparison\ can be translated into a Bessel
type differential equation for $\bar\beta$ :
\eqn\diffeqf{
\left[4{d\over dt}\left(t^{1-d/2}{d\over dt} t^2\right)
+t^{2-d/2}\right]\bar\beta(t)=0\,.}
If we define the function $g(t)$ via $\bar\beta(t)=t^{-2}g(t)$
 then $g(t)$ satisfies :
\eqn\diffeqg{
4t\, g''+4\left(1-{d\over2}\right)\,g'+g=0\,.}
The homogenous equation is supplemented with a matching condition to the
first two terms in the expansion.
The indicial equation  for \diffeqg, $r(r-d/2)=0$, is
quadratic showing that there
are two independent solutions. In  particular for $d$ an even dimension
the indices for the two solutions differ by an integer signaling that one
of the solutions contains a logarithm.

All these features can be seen explicitly
by writing down the general solution for $g_{ij}(x,\rho)$ following from
\diffeqf, \diffeqg:
\eqn\soldiffeqffb{\eqalign{
g_{ij}(\rho)=&\gz_{ij}+\rho\go_{ij}\cr
&\quad+\left\lbrace{2(d-2)\over \dal^2}
+{\rho\over\dal}+{c_1\over\dal^2}(\rho\dal)^{d/4}J_{-d/2}(\sqrt{\rho\dal})
+{c_2\over\dal^2}(\rho\dal)^{d/4}
J_{d/2}(\sqrt{\rho\dal})\right\rbrace X_{ij}}}
where
\eqn\BesselJ{
J_{\nu}(z)=\left({z\over 2}\right)^\nu\sum_{k=0}^\infty
{(-1)^k\over k!\Gamma(1+\nu+k)}\left({z\over2}\right)^{2k}}
and $\go_{ij}$ as given in \gone;
$c_1,c_2 $ are arbitrary coefficients. We choose the constant
$c_1$ such that when $c_2=0$ we recover
$g_{ij}(\rho)=\gz_{ij}+\rho \go_{ij}+{\cal O}(\rho^2)$. This gives
\eqn\cone{
c_1=-2(d-2)2^{-d/2}\Gamma(1-{d\over2})
=-{2(d-2)2^{-d/2}\pi\over\sin({\pi d\over2})
\Gamma({d\over2})}}
The arbitrariness of $c_2$ is the FG ambiguity: in an even
dimension $d=2n$, one has a series starting with $\rho^n$ whose
normalization is not fixed and simultaneously the first series
becomes singular. One can obtain a finite solution in an even
dimension using  the fact that the following limit exits
\eqn\BesselY{
\lim_{d\to 2n}{J_{d/2}(\sqrt{t})\cos\left({\pi d\over2}\right)
-J_{-d/2}(\sqrt{t})\over\sin\left({\pi d\over2}\right)}
\equiv Y_{n}(\sqrt{t})}
Then choosing
\eqn\ctwo{
c_2=-\cos\left({\pi d\over2}\right)c_1=
{2(d-2)\over 2^{d/2}\Gamma({d\over2})}\pi\cot\left({\pi d\over 2}\right)}
the expression \soldiffeqffb\
will have a unique well-defined limit in even dimension
which satisfies also the matching condition. The solution is nonlocal
since $Y_n$ has a logarithmic dependence of  $\rho\box$,
c.f. e.g. \AS.

We stress that this finite solution evaluated at $\rho=\bar\rho$
is not the renormalized expectation value of the
 energy momentum tensor: while it
has the correct leading logarithmic $\bar\rho$ -- independent term
it contains  also terms where the logarithm is multiplied by
powers of $\bar\rho$, i.e. inverse powers of $\mu^2$. The correct
procedure is to use the regulated $g_{ij}$ to evaluate the
action. The integration over $\rho$
should be extendable to infinity. The expression for  the effective
action in noninteger dimensions is therefore Weyl invariant the only possible
poles reflecting the  singularity at $0$ of the $\rho$-integration.
The counterterms needed will give now renormalized expressions which have
correct analyticity  as we reviewed
at the beginning of Section 2. An explicit verification of the analytic
structure in perturbation theory around flat space was done in \AF.
For a discussion of ``holographic renormalization'' see \dHSS\BFS\Skenderis.

\newsec {The Solution of the PBH equations for $\gth$ }

The results of the previous section prove  that the normalizations
of the cohomologically nontrivial Bach tensors which contain terms
with one curvature in all  $\gn$ are uniquely determined by the
PBH equations. In general there are more than one Bach tensors
contributing to a given $\gn$ whose expressions start with
$1,2,\dots,n$ curvature tensors. A complete proof that all the
normalizations are related requires a determination of the
relative contributions of these various tensors to $\gn$. We cannot offer a
general proof. Instead we will check that for the first non-trivial
case $\gth$ when three Bach tensors are present, their coefficients are
completely fixed.

In order to solve the PBH equation for $\gth$ we expand its Weyl variation
in terms of the lower $\gn$:
\eqn\vari{\eqalign{
\delta \gth_{ij}=&-4\sigma\gth_{ij}
-{1\over4}(\nabla_k\go_{ij})\go{}^{kl}\nabla_l\s
-{1\over4}\left[\go_{ik}\nabla_j\go{}^{kl}
+\go_{jk}\nabla_i\go{}^{kl}\right]\nabla_l\s\cr
\noalign{\vskip.2cm}
&+{1\over2}\left[\nabla_l\gt_{ij}
-{1\over3}\left(\nabla_i\gt_{jl}+\nabla_j\gt_{il}
\right)\right]\nabla^l\s
+{1\over6}\left(\nabla_i\go{}^2\!_{jl}
+\nabla_j\go{}^2\!_{il}\right)\nabla^l\s\cr
\noalign{\vskip.2cm}
&+{1\over3}\left(\gt_{jk}\na_i\na^k\s+\gt_{ik}\na_j\na^k\s\right)
-{1\over12}\left(\go{}^2\!_{il}\,\na_j\na^l\s+\go{}^2\!_{jl}\,
\na_i\na^l\s\right
)}}
where $\go$ and
$\gt$ are the  solutions of the PBH equations presented in
\ISTY. We remark that $\gt$ has two free parameters $c_1,c_2 $
which reflects the fact that there are two symmetric tensors
with four derivatives built from $\gz_{ij}$ which transform
homogenously with weight two.

On the other hand $ \gth_{ij}$ can be expanded in a general basis
of tensors with six derivatives:
$\box\nabla_i\nabla_j R$ , $\box^2 R_{ij}$,
$R\nabla_i\nabla_j R$, $R_{ij}\box R$, etc. A complete basis,
which contains 59 elements, can be found in \Fulling.
All these terms transform with weight four under constant
rescalings of the metric. Under local rescalings their transformations
contain up to six derivatives of $\sigma$. In contrast to this
\vari\ only contains at most two derivatives of $\sigma$.
Matching the coefficients we obtain the solution for $\gth $, unique up
to terms with six derivatives transforming homogenously. There are
eight such symmetric tensors.

We isolate in the solution the contribution of the three Bach tensors
${\cal B}^\a$  defined as (c.f.\typeBc)
\eqn\bach{  {\cal B}^\a_{ij}={1\over\sqrt{g}}{\delta\over\delta g^{ij}}
\int d^dx\sqrt{g}\,{\cal C}_\a}
for $\a=1,2,3$. The expressions  ${\cal C}_\a$ listed below
transform homogenously with
weight six under a Weyl transformation.
\eqn\wey{{\cal C}_1=C^{klmn}C_{mnpq}C^{pq}{}_{kl}}
\eqn\weyy{{\cal C}_2=C^{ijkl}C_i{}^m{}_k{}^n C_{jmln}}
\eqn\weyyy{{\cal C}_3=C^{ijkl}\box C_{ijkl} +\dots}
The complete expression for ${\cal C}_3$ can e.g. be found in \Erdmenger.
The solution for $\gth$ contains then the contributions:
\eqn\bachi{
\gth_{ij}={1\over 576(d-6)}\left( 7{\cal B}^{(1)}_{ij}+4{\cal B}^{(2)}_{ij}
+6{\cal B}^{(3)}_{ij}\right)+\hbox{finite terms}}
We remark that even though $\gt$ contains the arbitrary coefficients $c_1,c_2$
the Bach terms do not depend on them verifying our conjecture that all the type
B coefficients in all the terms $\gn$ are uniquely determined.

The cohomologically nontrivial type A contribution, being given by
a combination of homogenously transforming terms, is not determined
by the PBH equations as discussed in \ST. If we solve the
equations of motion for e.g. the simplest action \actiona\ we
obtain the additional pole term in \bachi\  ${{\cal J}_{ij} \over
576(d-6)}$  where ${\cal J}_{ij}$ is the tensor corresponding to
the Euler density in $d=6$, i.e.
${1\over d-6}{\cal J}_{ij}=A_{(3)ij}^E$, c.f. \typeAc.

\newsec{Anomalies for the General Quadratic Action}

All the type B terms in $\gn_{ij}$ are determined  as discussed in
Sections 3 and 4. Combining this with the action dependent linear
relation between $\gn_{ij}$ and the anomalies allows us to
calculate directly the anomalies from the universal terms in
$\gn_{ij}$.

For the type A anomaly we need to calculate the relevant action
dependent contribution in the equation of motion and then use the
same linear relation to the anomaly. Alternatively we can use the
universal, action dependent relation proven in \ISTY.

We exemplify this type of calculation for  the action \lag\
containing general quadratic  terms in the curvatures. The
condition for this action to admit an AdS solution is:
\eqn\cond{ \Lambda={1\over2}d(d-1)+(d-3)\left({\a\over2}d^2(d+1)+
 {\b\over2}d^2+\c d \right )>0 }
The general relation between the anomalies and $g_{ij}^{(n)}$ is given
for \lag\  by  \tonee .

For the trace anomalies in $d=4$ we need $\gt$ whose general
expression is \ST:
\eqn\twod{ \gt_{ij}=-{B_{ij}\over16(d-4)}
-{a\over8(d-4)}\left({1\over4}C^2 \gz_{ij}-C^2_{ij}\right)+{\rm finite}}
the type B being fixed while the type A has an action dependent
parameter $a$. In order to find $a$ one solves the equation of
motion for $\gt$ isolating the type A combination
\eqn\Deltagtwo{
\Delta\gt_{ij}
=b_1 C^2 \gz_{ij}+b_2 C^2_{ij}}
We obtain for $b_1 , b_2$
\eqn\ab{\eqalign{
b_1
&=-{1\over32}-{\c\over4(d-4)(1+40\a+8\b-4\c)}+\hbox{finite as $d\to 4$}\cr
b_2
&={1\over8}+{\c\over(d-4)(1+40\a+8\b-4\c)}+\hbox{finite as $d\to 4$}}}
This gives
\eqn\avalue{
a={1+40\a+8\b+4\c\over1+40\a+8\b-4\c}}
Combining with \tonee\ this gives an expression for the anomalies
\eqn\ano{\langle T^i_i\rangle=
-{1\over8}\left\lbrace(1+40\alpha+8\beta-4\gamma)C^2
-(1+40\alpha+8\beta+4\gamma)E_4\right\rbrace}
matching exactly the standard, ``bulk'' calculation \HS,\NO,\BNTone,\Tup.

The coefficient of $E_4$  matches also the general formula of
\ISTY, namely it is essentially the value of the action evaluated
on AdS space.

Another check can be made in $d=6$ for the simplest action
\actiona. From \ISTY\ we know that the normalization of the
type A anomaly in $d=2n$ dimensions is ${4n\over 2^{2n}(n!)^2}$
where the factor $4n=2d$ is $R-2\Lambda$ evaluated on AdS$_{2d+1}$.
For $d=6$ this gives ${1\over 192}E_6$.
Using \bachi, the normalization of the Euler term
$\gth_{ij}={{\cal J}_{ij} \over 576(d-6)}+\dots$ and the linear relation \tone\
which gives for $\gth$ a factor $-3$ we find perfect agreement
for type A. The type B follows from the singular terms displayed
in \bachi\ and agrees with the results of \HS.

\newsec{Discussion}

The use of  PBH equations allows to uncover the universal features
of trace anomalies in  CFTs which have a holographic dual:

(i) The relative
normalizations of the type B trace anomalies are completely fixed.

(ii) One is left with two action dependent overall normalizations
for the type A and for the type B. Even though these
normalizations are action dependent their dimensional dependence
is fixed relating theories in different dimensions which are dual
to the same gravitational lagrangian.

The features above appear in a classical context relating directly
the nontrivial cohomology of the solutions of the PBH equations to
the anomalies. A very much related manifestation of the same
structure is the FG ambiguity \FG.

Though the detailed structure of the equations is different, their
spirit is very similar to the relation between chiral anomalies
and the Chern-Simons lagrangians through the descent equations.
This is satisfactory since in supersymmetric theories chiral and
trace anomalies appear in the same supermultiplet.

Following this analogy the gravitational lagrangian corresponds to
the elliptic genus, the coefficients of the curvatures being the
analogues of the chiral matter representation dependent traces.

The question if the trace anomalies of every CFT can be
represented by a holographic gravitational  lagrangian is
completely open. In particular the field theoretical meaning of
the completely fixed ratios of type B anomalies is intriguing.

\medskip
{\bf Acknowledgement:} We thank D.Perini for his collaboration

\vskip1cm
\noindent{\bf Appendix: PBH transformations}

Since the PBH transformations play an essential r\^ole,
we will briefly review them; more details can be found in \ISTY.

Following \FG\ and \HS\ we write the bulk metric in the form
$$
ds^2=G_{\mu\nu}dx^\mu dx^\nu
={l^2\over4}\left({d\rho\over\rho}\right)+{1\over\rho}g_{ij}(x,\rho)dx^i
dx^j\eqno(A.1)
$$
where the conformal boundary is at $\rho=0$. We will set the
length scale $l=1$. The PBH transformations are those
bulk-diffeomorphisms which leave the form of (A.1) invariant. They
are parameterized by a scalar function $\sigma(x)$ and change
$g_{ij}(x,\rho)$ as \ISTY
$$
\delta
g_{ij}(x,\rho)=2\sigma(1-\rho\p_\rho)g_{ij}(x,\rho) +\nabla_i
a_j(x,\rho)+\nabla_j a_i(x,\rho)\eqno(A.2)
$$
where $a_i=g_{ij}a^j$ and
$$
a^j={1\over 2}\int_0^\rho d\rho'
g^{jk}(x,\rho')\p_k\s(x)\,.\eqno(A.3)
$$
The covariant derivatives in (A.2) are w.r.t. the metric $g_{ij}(x,\rho)$.

\listrefs

\bye